\documentclass[10pt,conference]{IEEEtran}
\usepackage{cite}
\usepackage{amsmath,amssymb,amsfonts}
\usepackage{algorithmic}
\usepackage{graphicx}
\usepackage{textcomp}
\usepackage{xcolor}
\usepackage[hyphens]{url}

\def\BibTeX{{\rm B\kern-.05em{\sc i\kern-.025em b}\kern-.08em
    T\kern-.1667em\lower.7ex\hbox{E}\kern-.125emX}}

\usepackage{amsmath,mathtools,amssymb}

\providecommand{\todo}[1]{{\protect\color{red}\noindent {\bf [TODO]}\emph{#1} {\bf [/TODO]}}}
\usepackage{titlesec}
\usepackage{multirow}
\usepackage{tikz}
\usepackage{adjustbox}
\titlespacing*{\section}
{0pt}{\parsep}{\parsep}
\titlespacing*{\subsection}
{0pt}{\parsep}{\parsep}
\usepackage[ruled,vlined,linesnumbered]{algorithm2e}

\newcommand\ket[1]{\left|#1\right\rangle}

\newcommand\circlenum[1]{\raisebox{.5pt}{\textcircled{\raisebox{-.9pt} {#1}}}}

\def\frameworkName {CollComm}
\def\frameworkNameSpace {CollComm }
\def\titleName {\frameworkName: Enabling Efficient Collective Quantum Communication Based on EPR buffering}

\author{
Anbang Wu\\
\texttt{anbang@ucsb.edu} \\
UC, Santa Barbara
\and
Yufei Ding\\
\texttt{yufeiding@cs.ucsb.edu} \\
UC, Santa Barbara
\and
Ang Li\\
\texttt{ang.li@pnnl.gov} \\
Pacific Northwest National Laboratory
}
\begin{document}

\title{\titleName}
\maketitle
\thispagestyle{plain}
\pagestyle{plain}


\begin{abstract}
The noisy and lengthy nature of quantum communication hinders the development of distributed quantum computing. The inefficient design of existing compilers for distributed quantum computing worsens the situation. Previous compilation frameworks couple communication hardware with the implementation of expensive remote gates. However, we discover that the efficiency of quantum communication, especially collective communication, can be significantly boosted by decoupling communication resources from remote operations, that is, the communication hardware would be used only for preparing remote entanglement, and the computational hardware, the component used to store program information, would be used for conducting remote gates. 
Based on the observation, we develop a compiler framework to optimize the collective communication happening in distributed quantum programs.
In this framework, we decouple the communication preparation process in communication hardware from the remote gates conducted in computational hardware by buffering EPR pairs generated by communication hardware in qubits of the computational hardware.
Experimental results show that the proposed framework can almost halve the communication cost of various distributed quantum programs, compared to the state-of-the-art compiler for distributed quantum computing.
\end{abstract}
\section{Introduction}


Distributed quantum computing (DQC) architecture is being actively explored~\cite{Laracuente2022ShortRangeMN, Jnane2022MulticoreQC} to scale up the monolithic near-term quantum processing unit (QPU) whose system size is small (up to a few hundred qubits) due to the fabrication limitation~\cite{Li2020TowardsES}. In a DQC system, multiple small QPUs (a.k.a compute nodes) are coordinated by inter-node communications to form a large quantum computing (QC) system. A general DQC system may interface with a hierarchical quantum network and contain several computing levels, e.g., the computing cluster driven by quantum chiplets~\cite{Laracuente2022ShortRangeMN} on short-range networks~\cite{Magnard2020MicrowaveQL, Gold2021EntanglementAS} and the quantum data-center including computing clusters linked by the long-range network~\cite{Monroe2014LargescaleMQ, Valivarthi2020TeleportationST, Rakonjac2021EntanglementBA}.

Inter-node communication is inevitable when executing programs on DQC hardware. Inter-node communication relies on EPR pairs generated by communication qubits (qubits specifically designed for EPR generation) and makes gates on inter-node data qubits (qubits used to store program information) executable.
Compared to in-node communication between data qubits, inter-node communication is far more error-prone and time-consuming~\cite{time-slice, Young2022AnAF} and should be carefully optimized when distributing quantum programs by DQC compilers.
Existing DQC compilers can be divided into two categories---
either ignore the low-level implementation of quantum communication and perform program optimizations on the logical level~\cite{Beals2013EfficientDQ, Moghadam2016OptimizingTC, AndresMartinez2019AutomatedDO, Davarzani2020ADP, daei2020optimized, Dadkhah2022ReorderingAP} (equivalent to assuming an unbounded number of communication qubits)
or consider a limited number of communication qubits due to the hardware fabrication difficulty~\cite{Ferrari2021CompilerDF, Diadamo2021DistributedQC, autocomm}. 
Works in the former category focus on distributing circuits to compute nodes as the parallelism between inter-node communication and the feasibility of executing any inter-node circuit blocks are naturally enabled by the unbounded communication qubits.
Thus it is hard to guarantee the performance of works in the first category when applied to DQC hardware in the near future where limited communication resource is expected on compute nodes. 

Works in the latter category target optimizing communication on DQC hardware with limited communication qubits but fall short in implementing multi-node operations  (i.e., collective communication). For example, if the four qubits of a CCCX gate are evenly distributed on four compute nodes, with the few ($\le 2$) communication qubits considered in~\cite{Ferrari2021CompilerDF, Diadamo2021DistributedQC, autocomm}, we cannot implement the CCCX gate \textit{collectively} by sharing all the three control qubits to communication qubits that are in the same node as the target qubit and then executing the CCCX gate locally. Actually, for works in the second category to implement multi-node  gates (or unitary blocks), they need to decompose the target operation into smaller unitary blocks or basis gates (e.g., CX+U3~\cite{Qiskit}), which would inevitably incur higher communication costs than implementing the multi-node operation collectively. 
Besides the inefficiency in performing collective communication, these works admit low communication throughput. They usually perform inter-node communications with quite limited parallelism and even sequentially.


\begin{center}
    \textit{Thus, can we overcome the inefficiency \\cast by limited communication qubits while \\respecting the constraints on DQC hardware?}
\end{center}

We observe that it's the inefficient use of communication qubits rather than the amount of them
that results in the unpromising communication performance of works in the second category. In these works, communication qubits are coupled with inter-node operations that we cannot use them to prepare new EPR pairs for the next communication when they are being used for implementing inter-node operations, let alone to accommodate the multi-node gates and concurrent inter-node communication.
\textit{Our answer to the above question is thus} YES \textit{with our key insight}:

\begin{center}
\textit{Rather than coupling inter-node operations \\ with communication qubits as in previous works, \\
decoupling them gives more benefits for communication.}
\end{center}

\begin{figure}[t]
    \centering
    \includegraphics[width=0.42\textwidth]{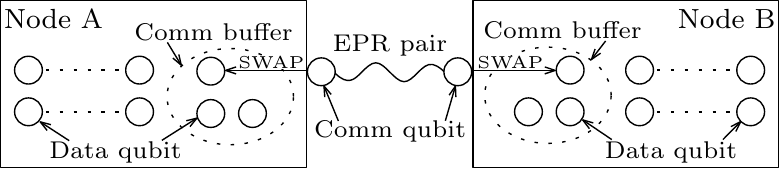}
    \caption{The communication buffer which buffers the EPR pair produced by communication qubits.}
    \label{fig:combufintro}
\end{figure}

Specifically, we invent a new versatile computing component called \textit{communication buffer} based on idle data qubits of compute nodes to decouple inter-node operations from communication qubits. In detail, communication qubits are always used for preparing EPR pairs and once an EPR pair is generated, we would swap it from communication qubits to data qubits in the communication buffer, as shown in Figure~\ref{fig:combufintro}. In this way, we can use the EPR pairs stored in the communication buffer to conduct inter-node operations. The communication buffer essentially provides an abstraction or intermediate layer that is able to approximate the ideal DQC hardware (the one with unlimited communication qubits) on near-term DQC hardware which is expected to have few communication qubits.
As long as the communication buffer is large enough, we can implement multi-qubit inter-node operations collectively and admit a large number of concurrent inter-node communication requests. On the other hand, since the communication buffer exploits idle data qubits in DQC hardware, it also boosts the device utilization rate.

\begin{figure}[t]
    \centering
    \includegraphics[width=0.48\textwidth]{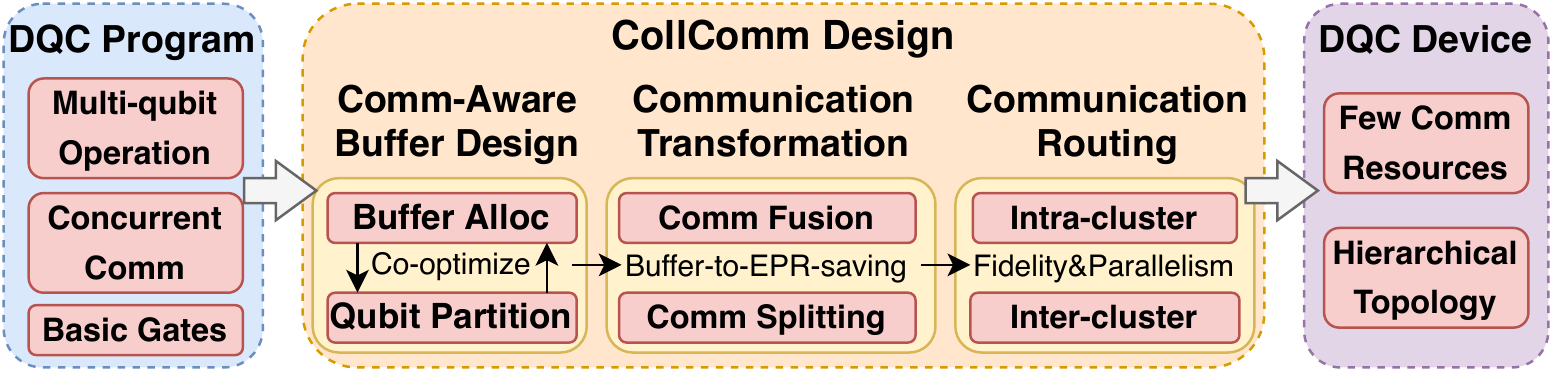}
    \caption{The design overview of {\frameworkName}. 
    }
    \label{fig:designoverview}
\end{figure}

Based on the proposed communication buffer, we further develop the first buffer-based communication optimization framework, \textit{\frameworkName} as shown in Figure~\ref{fig:designoverview}. {\frameworkName} provides excellent optimizations for collective communication which is unexplored by existing DQC compilers. {\frameworkName} consists of three key compiler passes. Firstly, given a qubit mapping, the first pass would inspect the communication characteristics (e.g., the maximal number of concurrent inter-node communication and the largest inter-node circuit block) of each compute node, and then produce the communication buffer allocation that potentially enables maximal throughput and fastest EPR generation rate on each node. As in Figure~\ref{fig:designoverview}, the buffer allocation would be co-optimized with the qubit mapper to achieve a good balance in the number between data qubits and the qubits in the buffer. 
With the well-tuned communication buffer, the second pass performs buffer-based communication transformations to reduce the EPR pair consumption. This pass would try to fuse multi-node circuit blocks as decomposed operations usually incur higher communication resource consumption than implemented collectively.
This pass also explores several optimizations which dynamically convert qubits in the communication buffer back to normal data qubits to reduce the communication cost.
Finally, the communication routing pass exploits the communication buffer to boost the communication throughput while reducing the communication cost induced by the hierarchical DQC network. This pass tries to squeeze out all EPR pairs in the communication buffer for concurrent inter-node communication. 
This pass further reduces hardware-induced communication costs by smartly relaying communications between distant (intra-cluster and inter-cluster) nodes.


Our contributions are summarized as follows:
\begin{itemize}
    \item We invent the communication buffer which provides an abstraction to approximate ideal DQC hardware and reveals vast communication optimization opportunities.
    \item We propose an efficient communication buffer allocation pass by inspecting  the communication characteristics of the distributed quantum program. This pass lays the foundation for communication optimizations. 
    \item We propose two buffer-based communication optimization passes which reduce both program-intrinsic and hardware-induced communication costs remarkably.
    \item  Compared to the state-of-the-art baseline~\cite{autocomm}, \frameworkNameSpace significantly reduces the inter-cluster communication resource consumption and the latency of various programs by 50.4\% and 47.6\% on average, respectively. 

\end{itemize}


\section{Background}\label{sect:bg}

In this section, we only introduce the essential background knowledge of distributed quantum computing (DQC). We recommend~\cite{nielsen2002quantum} for the basics of quantum computing. Without ambiguity, quantum/remote communication in the following sections specifically refers to inter-node communication. And for simplicity, we use \textit{block} to denote a series of gates.





\paragraph{Quantum entanglement and DQC:} 
Quantum communication relies on the remote EPR (Einstein–Podolsky–Rosen) entanglement in a pair of qubits that belongs to different quantum nodes. An EPR entangled qubit pair (a.k.a an EPR pair) holds the entangled two-qubit state $\frac{\ket{00}+\ket{11}}{\sqrt{2}}$. Other commonly-used entangled states in DQC include the GHZ state $\frac{\ket{000}+\ket{111}}{\sqrt{2}}$ and the cat-state $\frac{\ket{0}^{\otimes n}+\ket{1}^{\otimes n}}{\sqrt{2}}$, a generalization of the GHZ state on $n$ qubits. 

The remote EPR pair between two nodes can be generated, e.g., via microwave~\cite{Magnard2020MicrowaveQL}, or by interfering photons emitted from separate nodes~\cite{Monroe2014LargescaleMQ}. The former way is used to create short-range interconnect networks for quantum chiplets~\cite{Laracuente2022ShortRangeMN}, while the latter way may achieve long-range quantum networks like the quantum internet~\cite{Wehner2018QuantumIA}. 
The Fidelity of EPR pairs on the short-range network is higher: the on-chip quantum link/EPR pair between nodes can achieve a fidelity of at most 98.6\%~\cite{Gold2021EntanglementAS} while the fidelity of a long-range quantum link is at most 90\% in the recent studies~\cite{Valivarthi2020TeleportationST, Rakonjac2021EntanglementBA}. The preparation of a long-range quantum link is also much more time-consuming than preparing a short-range link.
Similar to classical distributed computing, quantum compute nodes can also form a hierarchical network topology where compute nodes connected by a short-range network form a computing cluster, and clusters are inter-connected by the long-range network.

\begin{figure}[t]
    \centering
    \includegraphics[width=0.24\textwidth]{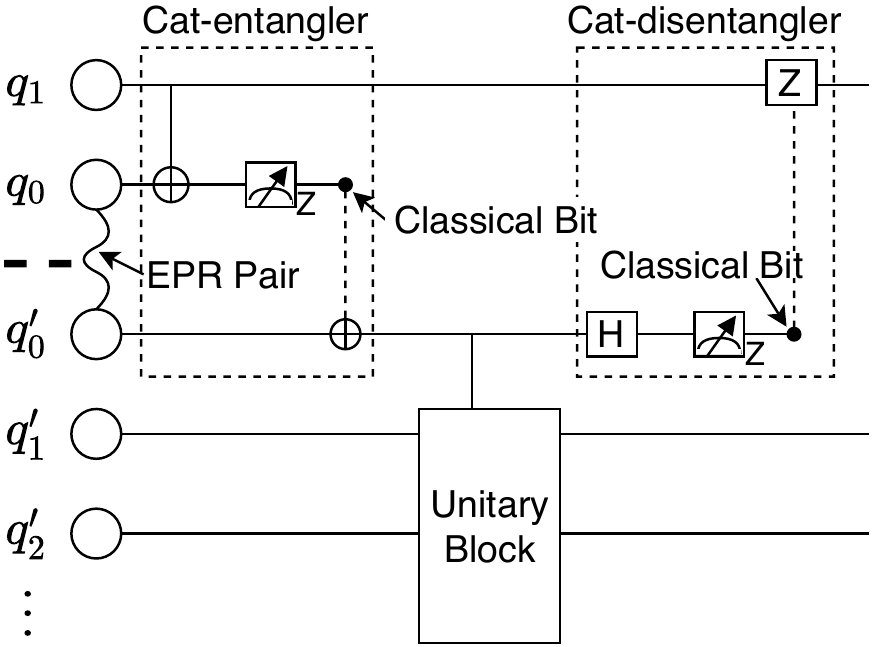}\hfill{}\includegraphics[width=0.24\textwidth]{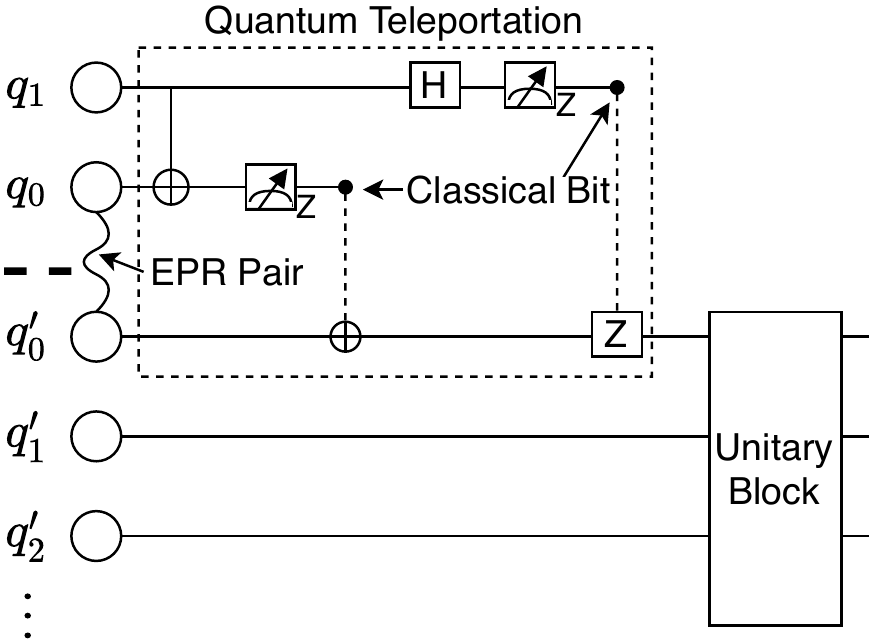} \\
    {\small \hphantom{} \hspace{10pt} (a) \hspace{100pt} (b)}
    \caption{(a) The Cat-Comm protocol. (b) The TP-Comm protocol. For the gate (e.g. the final $Z$ gate) conditioned on the Z measurement, it is executed only if the post-measured state is $\ket{1}$, corresponding to the classical bit 1.}
    \label{fig:catcomm}
\end{figure}

Not all physical qubits on a DQC node can be directly used to establish the remote EPR entanglement with another node~\cite{AndresMartinez2019AutomatedDO}. Qubits able to construct remote EPR pairs are called \textit{communication qubits}~\cite{Caleffi2020TheRO} and serve as the base of quantum communication. To distinguish the communication qubits from other qubits in the node, we would refer to other qubits that are not designed for quantum communication as \textit{data qubits} or \textit{computational qubits}. For example, in Figure~\ref{fig:catcomm}, $q_0$ and $q_0'$ are communication qubits while $q_1$ and $q_1'$ are data qubits. In the programming aspect, data qubits used to store the information of the quantum program are also called \textit{program qubits}.

\paragraph{Quantum communication protocols:} The quantum no-clone theorem~\cite{nielsen2002quantum} makes quantum data not replicable between compute nodes. 
To overcome this limitation, quantum communication protocols rely on EPR pairs for data sharing. 
There are mainly two communication protocols: the Cat-Comm shown in Figure~\ref{fig:catcomm}(a) and the TP-Comm shown in Figure~\ref{fig:catcomm}(b). In Figure~\ref{fig:catcomm}(a),
Cat-Comm utilizes the cat-entangler and -disentangler to first share the state of $q_1$ to $q_0'$, perform the target controlled-unitary block and then revoke the sharing. 
Cat-Comm is used to implement many commonly-used multi-node/collective operations~\cite{Hner2021DistributedQC}, e.g., broadcast (sharing one qubit to all other compute nodes), multicast (sharing one qubit to multiple other compute nodes), and reduce (equivalent to broadcast up to local H gates). TP-Comm, on the other hand, exploits quantum teleportation~\cite{nielsen2002quantum} to move qubits between compute nodes. For the TP-Comm in Figure~\ref{fig:catcomm}(b), it first moves $q_1$ to $q_0'$ and then performs the target unitary block. The differences between Cat-Comm and TP-Comm come from two aspects: Cat-Comm can only be used to implement remote controlled-unitary blocks while TP-Comm can implement any inter-node unitary blocks; After Cat-Comm, the qubit being shared (i.e., $q_1$ in Figure~\ref{fig:catcomm}(a)) will recover its state while TP-Comm will leave the teleported state at the new qubit ($q_0'$ in Figure~\ref{fig:catcomm}(b)) and destroy the original qubit ($q_1$). In a distributed quantum program, both TP-Comm and Cat-Comm may appear in (be a part of) a general collective communication (the communication involving multiple nodes). To avoid ambiguity, in the following sections, when we say \textit{sharing} one qubit to another node, we mean using Cat-Comm; when we say \textit{moving} or \textit{teleporting} one qubit to another node, we mean using TP-Comm.





\section{Problem and Framework Design}

\subsection{Collective Communication in DQC}\label{sect:problem}

\begin{figure}[h]
    \centering
    \includegraphics[height=0.17\textwidth]{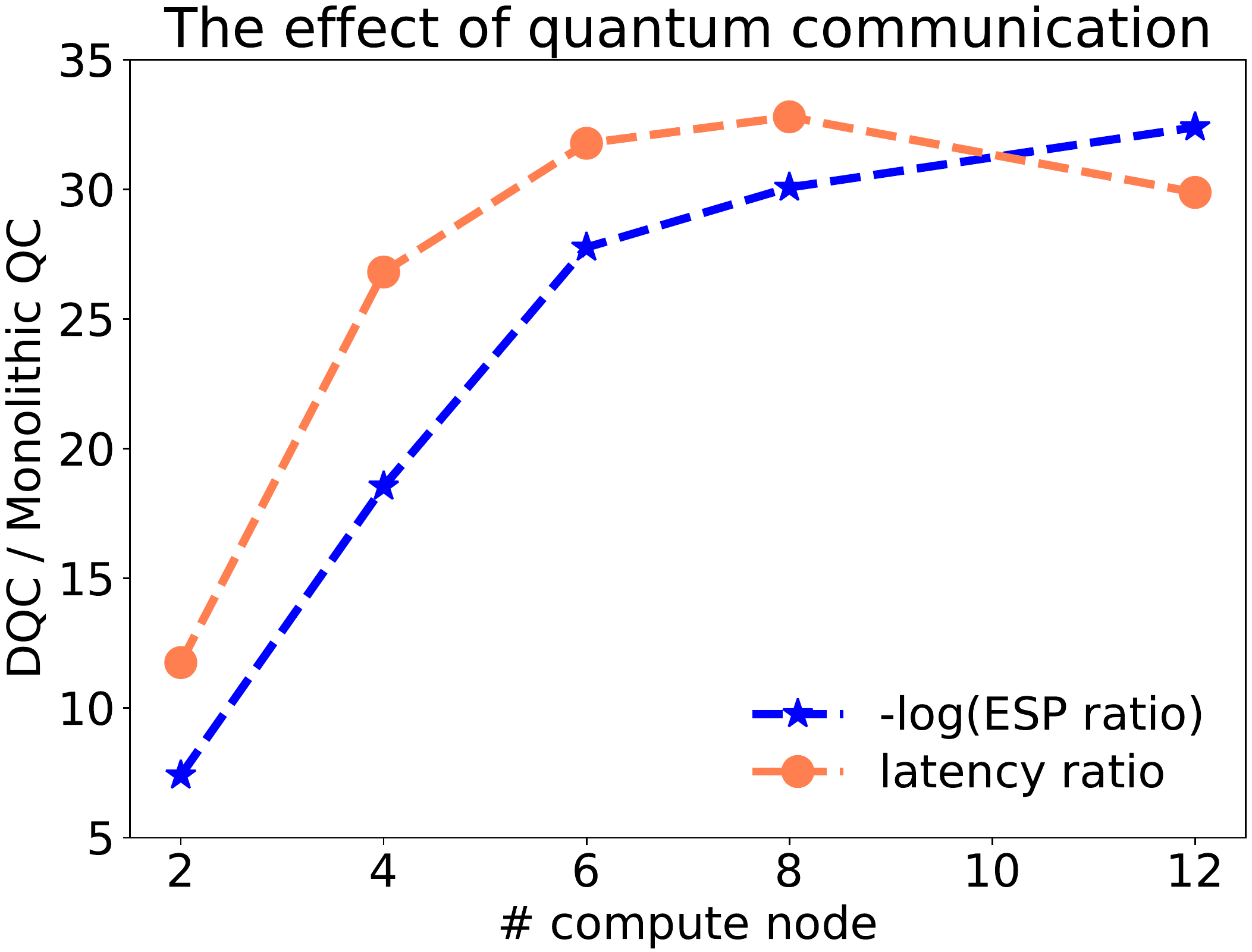} \hfill{} \includegraphics[height=0.17\textwidth]{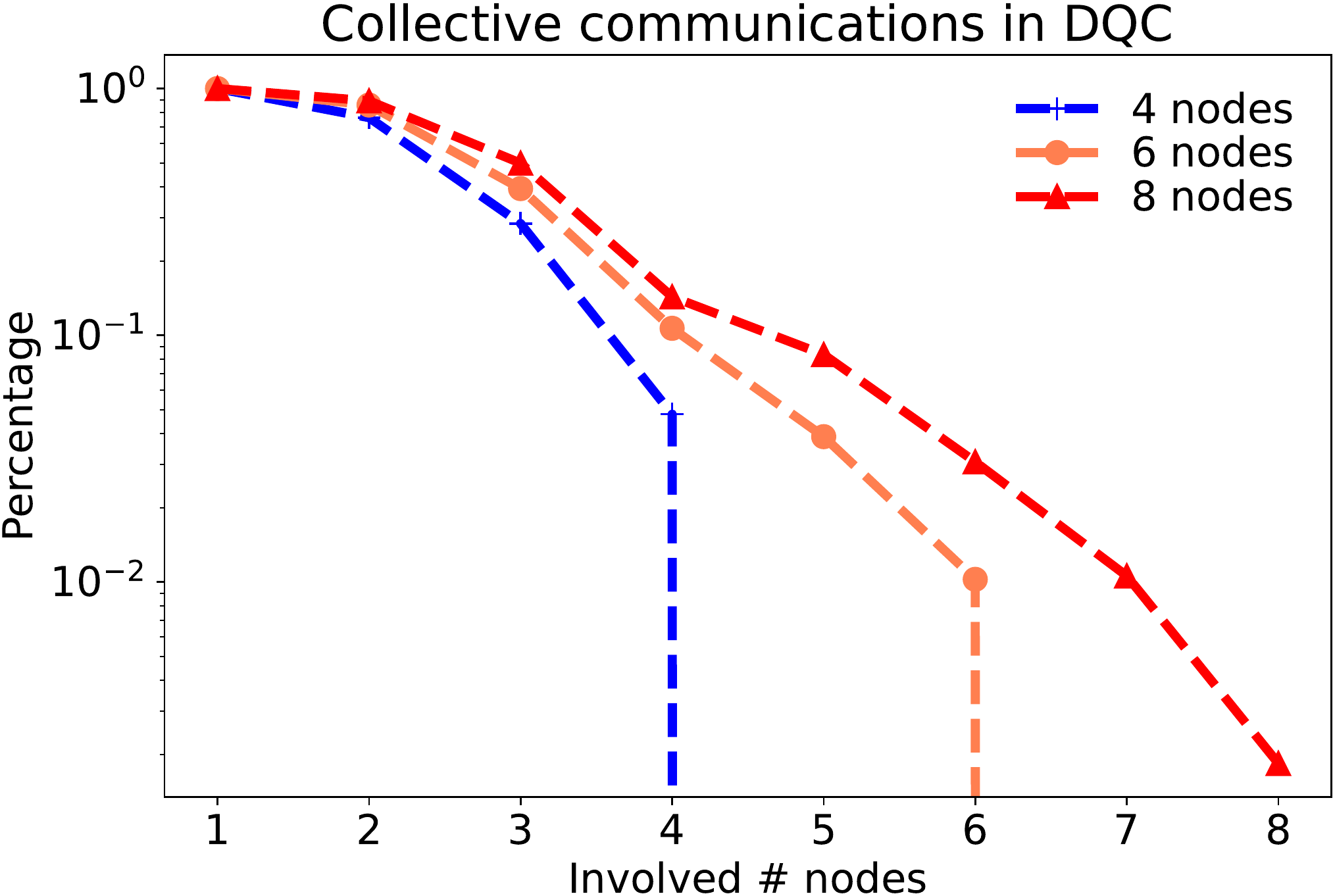}\\
    {\small \hphantom{} \hspace{10pt} (a) \hspace{116pt} (b)}
    \caption{(a) The comparison of DQC and monolithic QC by applying Qiskit~\cite{Qiskit} to the QFT program. Gate fidelity and latency are derived from~\cite{isailovic2006interconnection, Diadamo2021DistributedQC, Gold2021EntanglementAS} and IBM's device data. The time for preparing the EPR pair is ignored.
    (b) The average percentages of multi-qubit gates from various quantum circuits that involve more than a certain number of nodes, using the qubit-node mapping by METIS ~\cite{Karypis1998AFA}. Circuits are collected from ~\cite{Qiskit, revlib}. }
    \label{fig:effectcomm}
\end{figure}

Like classical distributed computing, quantum communication bottlenecks the performance of DQC, both in fidelity and latency. 
Figure~\ref{fig:effectcomm}(a) shows the comparison of the DQC version quantum Fourier transformation (QFT) and the monolithic QC version QFT (both compiled by Qiskit~\cite{Qiskit} with remote communications treated in the same way as local communications).
As shown in Figure~\ref{fig:effectcomm}(a), the ESP (Estimated Probability of Success) and latency of programs run on DQC hardware will be dozens of times worse than the ones executed on a monolithic QC device.
The considerable performance discrepancy in Figure~\ref{fig:effectcomm}(a) results not only from the noisy and tedious nature of quantum communication~\cite{nielsen2002quantum} but also from the bad adaptability of monolithic QC compilers in addressing inter-node communication overheads. While the hardware properties of quantum communication are hard to improve significantly in the near future, the software compilation of distributed quantum programs should try its best to mitigate the detrimental effects of quantum communication. Efficient compilation support is critical to making DQC a feasible solution to overcome the awful scalability of monolithic quantum devices~\cite{Laracuente2022ShortRangeMN, Li2020TowardsES}.

We observe that the efficient implementation of \textit{collective communication} (the communication involving multiple nodes) in distributed quantum programs is critical to promoting DQC's computational potential. 
By collecting statistics on various quantum circuits (QFT, arithmetic functions, encoding circuits, etc.) from existing quantum benchmarks~\cite{Qiskit, revlib}, we observe that 77.6\%, 20.2\%, and 10.6\% of quantum gates involve more than 3, 6, and 9 qubits, respectively, on average. This demonstrates the potential existence of abundant collective communications when executing these circuits on DQC hardware. 
Figure~\ref{fig:effectcomm}(b) shows the statistics of collective communication in distributed quantum circuits, with the METIS algorithm~\cite{Karypis1998AFA} to map qubits to compute nodes. As shown in Figure~\ref{fig:effectcomm}(b), there are about 28.4\% of multi-qubit gates require communications between more than 3 nodes, when running collected circuits on 4 compute nodes. The percentage grows up when we run these circuits on 8 compute nodes, where 49.8\% of multi-qubit gates involve computation on more than 3 nodes. Therefore, to promote the efficiency of DQC, we must optimize the implementation of collective communication wisely, e.g., reducing the communication latency and the number of used EPR pairs.

\subsection{Communication Buffer and Framework Design}

\begin{figure}
    \centering
    \includegraphics[width=0.38\textwidth]{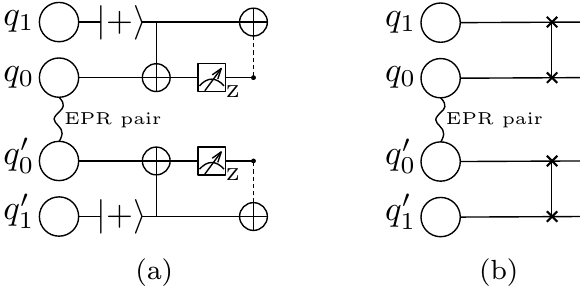}
    \caption{The buffering protocols which transfer the remote EPR pair on communication qubits ($q_0$ and $q_0'$) to data qubits ($q_1$ and $q_1'$). 
    }
    \label{fig:virtualcomm}
\end{figure}

The optimization of collective communication still remains unexplored by current DQC compilers. Due to the shortage of communication qubits in near-term DQC hardware, it is often impractical to execute one multi-node operation \textit{collectively}, i.e., teleport/share all involved qubits to one node and then implement the target operation locally, let alone explore specialized circuit transformation and routing strategies for optimizing the cost of collective communication. For example, if the three qubits of a Toffoli gate are distributed in three fully-occupied nodes and each node only has one communication qubit, then this remote Toffoli gate cannot be executed unless it is decomposed into a series of CX gates and single-qubit gates with the cost of consuming more EPR pairs. 
Thus to enable efficient collective communication in DQC, we need to address challenges coming from both hardware incapability and software inefficiency.

We observe that idle data qubits can be used to store remote EPR pairs generated by the communication qubit, as indicated by buffering protocols in Figure~\ref{fig:virtualcomm}. We call data qubits with remote EPR pairs stored \textit{virtual communication qubits}. Virtual communication qubits can also be  used for implementing inter-node operations. That's to say, 
by buffering protocols in Figure~\ref{fig:virtualcomm}, one physical communication qubit can emulate many virtual communication qubits, thus overcoming the hardware incapability. Virtual communication qubits constitute a software-defined communication facility and we refer to it as a \textit{communication buffer}. The communication buffer provides an abstraction layer that decouples the remote EPR entanglement preparation process from the execution of distributed quantum programs. A communication buffer can hold any number of remote EPR pairs as long as there are sufficient data qubits. 
The emergence of the communication buffer defines a new communication model for DQC: the physical communication qubit serves as the factory of virtual communication qubits and the communication buffer serves as the component for conducting inter-node communication. Compared to the communication model adopted by existing DQC compilers where communication qubits are directly involved in the implementation of remote operations, the communication buffer-based communication model can overcome the shortage of `physical communication qubits' and pave the way for optimizing collective communication.

Equipped with the communication buffer, we are now able to address the software inefficiency in optimizing collective communication.
Specifically, we design a buffer-based framework with three key passes, for an end-to-end compiler optimization for collective communications. 

Firstly, we design a communication-aware buffer allocations pass in Section~\ref{sec:pipeline} which determines the size of each node's communication buffer according to the communication characteristics (e.g., the maximum communication throughout and the longest remote operation) of the distributed quantum circuit. 
The design insight is that we can hide the latency of EPR preparation behind the execution of remote operations. A well-tuned communication buffer should enable a large EPR generation rate (thus high communication throughput) and not occupy too many data qubits which could alternatively be used as program qubits (to potentially achieve less node occupancy and fewer inter-node communication).
This pass utilizes high-level communication characteristics to produce a good answer to the trade-off between virtual communication qubits and program qubits (data qubits that stores program information).

Secondly, we design a buffer-based communication transformation pass in Section~\ref{sec:commtransform}. This pass would fuse remote unitary blocks to form a large multi-node block (a.k.a collective communication block) in a greedy way (as long as fewer EPR pairs are consumed). The insight behind such a design is that the EPR pair consumption would likely decrease when implementing remote operations collectively. On the other hand, for a better trade-off between the virtual communication qubits and program qubits on each node, this pass would dynamically convert a virtual communication qubit to a program qubit as long as fewer EPR pair consumption is expected.

Finally, we design a communication routing pass that efficiently arranges remote communications on a hierarchical DQC system with two network levels where we need to choose the communication path for communication between distant nodes.
The design of this pass is based on the observation that shortest path-based strategies for routing two-node or in-node communication cannot lead to an optimal footprint of collective communication. 
One insight in this pass is that Cat-Comm-based collective communication could be optimized by the minimum spanning tree of involved nodes. Also, there is a trade-off between parallelism and the footprint size of collective communications, especially for inter-cluster communication where EPR pairs are hard to prepare. For such a case, higher parallelism may be more favorable than always sticking to the minimal footprint of the collective communication.

\section{
    Buffer-based Communication Optimizations
}\label{sec:commopt}


In this section, we introduce the buffer-based communication optimizations: communication transformation to reduce the EPR pairs resulting from distributed computing and communication routing to minimize the communication overhead caused by the non-fully connected DQC network topology.

\subsection{Communication Transformation}
\label{sec:commtransform}

\begin{figure}[h]
    \centering
    \includegraphics[width=0.35\textwidth]{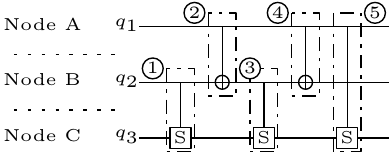}
    \caption{One example circuit extracted from quantum arithmetic circuits~\cite{Qiskit}, which is distributed over three compute nodes, with each node having one communication qubit. }
    \label{fig:examcirc}
\end{figure}

The basic benefit of the communication buffer is to preallocate EPR pairs before implementing inter-node operations, similar to the persistent communication in MPI~\cite{clarke1994mpi}. The communication buffer can also be used to store the teleported qubits in TP-Comm-based remote operations. In this way, we can perform lazy teleportation that we do not need to move the teleported qubit immediately back to its original nodes but leave it in the communication buffer (equivalent to converting one virtual communication qubit into a program qubit) since in the communication buffer, the teleported qubit cannot stall the EPR preparation process.

Besides these basic usages of the communication buffer, there are more advanced communication optimizations with communication buffers, as introduced in the following paragraphs. The communication transformation pass is a loop in which each iteration would run all these communication optimizations one by one. This loop continues until no more improvement on EPR pair consumption is reported.

\paragraph{Communication Fusion}
Collective communication, if implemented efficiently, may consume fewer EPR pairs than decomposed two-node communications.
Unfortunately, collective communication is not well optimized by state-of-the-art DQC compilers~\cite{Ferrari2021CompilerDF,Diadamo2021DistributedQC,autocomm} due to the lack of communication qubits. Using the virtual communication qubits in the communication buffer, we can reduce the EPR pair consumption by exploiting the potential opportunity for collective communication. As an example, we consider the distributed circuit shown in Figure~\ref{fig:examcirc}. To implement this circuit on the DQC hardware in which each compute node only has one communication qubit, five EPR pairs are needed with existing DQC compilers, e.g.,~\cite{autocomm}. However, with a communication buffer that occupies at least two qubits in node C, we can implement all two-node gates in Figure~\ref{fig:examcirc} together, requiring only need two EPR pairs to share $q_1$ and teleport $q_2$ to the communication buffer of node C, and then executing the circuit in Figure~\ref{fig:examcirc} locally in node C. The optimization described, which combines several inter-node operations into a collective communication block and then executes the collective block by collecting involved qubits into the same node with the help of the communication buffer, is called \textit{communication fusion}. By fusing remote unitary blocks with the communication buffer, we can significantly reduce the EPR pairs required to implement these remote blocks.

We adopt a greedy strategy to discover and utilize communication fusion opportunities. Given a preprocessed distributed circuit (we assume the circuit is preprocessed by \cite{autocomm}, i.e., two-node gates that can be implemented by one EPR pair are already aggregated into a two-node block), for example the one in Figure~\ref{fig:examcirc} in which each circled number denotes one two-node block, we describe the greedy strategy for the communication fusion starting from the two-node block \circlenum{1} as follows:
\begin{enumerate}
    \item Check the following two-qubit block \circlenum{2}. If \circlenum{2} and \circlenum{1} shares qubits, we proceed to Step 2. Otherwise,  \circlenum{2} commutes with \circlenum{1}. We exchange the position of these two blocks and restart from Step 1.\\
    -- In this example, since block \circlenum{2} and block \circlenum{1} share qubit $q_2$, block \circlenum{2} is a potential block for fusion and we proceed to Step 2.   
    \item Compute the EPR pair cost of executing block \circlenum{1}, \circlenum{2} collectively, i.e., by communication fusion. 
    \item Check whether the computed EPR pair cost in Step 2  surpasses the EPR pair cost of executing these blocks independently or not. If does not surpass, we fuse block \circlenum{1} and block \circlenum{2} and use the fused block as the new starting point of communication fusion. Otherwise, we use block \circlenum{2} as the new starting point. \\
    -- For the example circuit in Figure~\ref{fig:examcirc}, the computed EPR pair cost in Step 2 is 2 and does not surpass the EPR pair cost of executing them independently. Thus, for this example, we will fuse block \circlenum{1} and block \circlenum{2}.
    \item Repeat this process until no more fusion can be done.
\end{enumerate}

It's easy to see that after performing the above greedy procedure on the example circuit in Figure~\ref{fig:examcirc}, we can successfully fuse block \circlenum{1}--\circlenum{5} into a collective communication block that can be implemented with only two EPR pairs. In the communication transformation pass, we would invoke the greedy communication fusion procedure several times to squeeze out opportunities for collective communication in the underlying distributed program.

\paragraph{Communication Splitting}

\begin{figure}[t]
    \centering
    \includegraphics[width=0.41\textwidth]{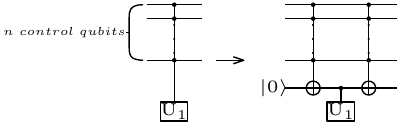}
    \caption{The circuit identity for splitting the multi-controlled unitary block with one ancillary qubit.}
    \label{fig:commsplit}
\end{figure}

We need to split the collective communication (multi-node unitary block) when one compute node is not large enough to accommodate all involved qubits. For example, we consider the multi-controlled-unitary block. If control qubits of the multi-controlled-unitary block are scattered in compute nodes and the number of them is larger than the node size, we cannot directly implement the related collective communication. 
One simple way to overcome this issue is to decompose the multi-controlled-unitary block into basis gates as in existing DQC compilers~\cite{Ferrari2021CompilerDF, autocomm}. However, this method would incur overwhelming communication costs. Even for a four-node CCCX gate, the decomposed block would require at least 11 EPR pairs for implementation. 
\begin{figure}[t]
    \centering
    \includegraphics[width=0.48\textwidth]{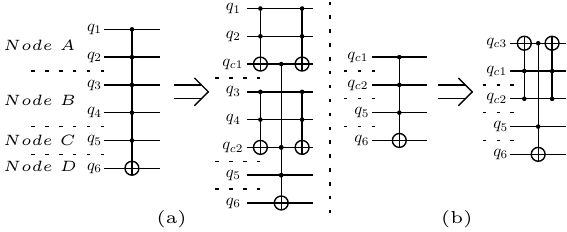}
    \caption{Recursive steps (i.e., (a)\&(b)) of splitting a four-node CCCCCX gate.}
    \label{fig:ccsplit}
\end{figure}

To avoid the heavy communication cost of the simple decomposition method, we design an efficient communication splitting method by utilizing the circuit identity in Figure~\ref{fig:commsplit}. Our communication splitting method first preserves qubits in the communication buffer of involved compute nodes, and then recursively applies the circuit identity in Figure~\ref{fig:commsplit} to split the multi-controlled-unitary block. As an example, we can split the CCCCCX gate in Figure~\ref{fig:ccsplit} as follows,
\begin{enumerate}
    \item For the node with $\ge 2$ control qubits (e.g., node A), use one qubit of the buffer (e.g., $q_{c1}$) to act as the ancilla in Figure~\ref{fig:commsplit}, as shown in Figure~\ref{fig:ccsplit}(a); 
    \item For the node with only control qubit, combining this control qubit with other control qubits to apply the splitting in Figure~\ref{fig:commsplit}, like the $q_{c1}$ and $q_{c2}$ in Figure~\ref{fig:ccsplit}(b).
    \\-- In this example, a four-node CCCX gate at most uses 6 EPR pairs (each Toffoli in Figure~\ref{fig:ccsplit}(b) uses two EPR pairs), almost half of that by trivial decomposition.
\end{enumerate}
Note that the above recursive splitting procedure should be stopped once we find the resulted communications are all executable, in order to exploit the benefit of large collective communication. A multi-controlled unitary of a size large than the buffer size could also be  executed collectively as we can use the communication buffer to swap qubits between nodes.

To some extent, our communication splitting optimization is based on the dynamical conversion between virtual communication qubits and program qubits (see $q_{c1}$ in Figure~\ref{fig:ccsplit}).
Our communication splitting method is generally applicable as any collective communication block is essentially a group of multi-controlled-unitary blocks~\cite{Aharonov2003ASP}.

\subsection{Communication Routing}
\label{sec:commroute}

\begin{figure}[t]
    \centering
    \includegraphics[width=0.48\textwidth]{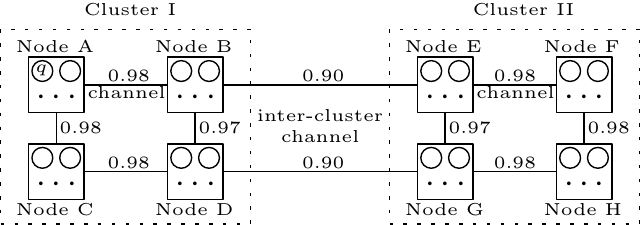}
    \caption{A hierarchical DQC system, which consists of two network levels: compute nodes forming clusters representing short-range networks and clusters forming a long-range quantum network. One communication channel denotes one EPR pair. Numbers on edges denote the fidelity of the related EPR pairs.}
    \label{fig:netmodel}
\end{figure}

\begin{table}[t]
    \caption{Quantitative data of operations in DQC (from~\cite{isailovic2006interconnection, Valivarthi2020TeleportationST, Rakonjac2021EntanglementBA, correa2018ultra}, with latencies normalized to CX counts). Fidelity is the averaged data and may not be uniform in the DQC system.} 
    \label{tab:netdata}
    \centering
    \footnotesize
    \renewcommand*{\arraystretch}{1.01}
    \begin{tabular}{|l|c|l|l|}
    \hline
    Operation & Latency & Fidelity \\ \hline
    Single-qubit gates & $t_{1q}$ $\sim$ 0.1 CX & $f_{1q} \approx$ 99.99\% \\ \hline
    CX and CZ gates & $t_{2q} =$ 1 CX & $f_{2q} \approx$ 99.80\% \\ \hline
    Measure & $t_{ms}$ $\sim$ 5 CX  & $f_{ms} \approx$ 99.60\% \\ \hline
    Intra-cluster EPR preparation 
    & $t_{iep}$ $\sim$ 12 CX & $f_{iep} \approx$ 98\%  \\ \hline
    Intra-cluster classical comm & $t_{icb}$ $\sim$ 1 CX & $\qquad$--- \\ \hline
    Inter-cluster EPR preparation 
    & $t_{oep}$ $\sim 1000$ CX & $f_{oep} \approx$ 90\%  \\ \hline
    Inter-cluster classical comm & $t_{ocb}$ $\sim 100$ CX & $\qquad$--- \\ \hline
    \end{tabular}
\end{table}



In this section, we consider the hierarchical DQC system shown in Figure~\ref{fig:netmodel} for demonstrating the buffer-based routing of collective communication. 
The quantitative data of operations in the hierarchical DQC system are summarized in Table~\ref{tab:netdata}.
When routing collective communication operations on the DQC computing system in Figure~\ref{fig:netmodel}, extra nodes are needed for relaying the communication as the EPR pairs between any two nodes are not always readily available in the communication buffer, either consumed or not able to be established directly. For example, sharing qubits in node A to node D may 
need both the EPR pairs between nodes A, B and nodes B, D, i.e., using node B for communication relaying.
More nodes in the communication (relay) path would incur higher communication overhead, w.r.t fidelity and latency.

Overall, we expect to minimize the extra communication overhead induced by the non-fully connected DQC network topology.
Specifically, we focus on optimizing the routing of Cat-Comm-based collective communication as each TP-Comm usually involves two nodes and can be well optimized by using the shortest communication paths between involved nodes.

\begin{figure}[t]
    \centering
    \includegraphics[width=0.40\textwidth]{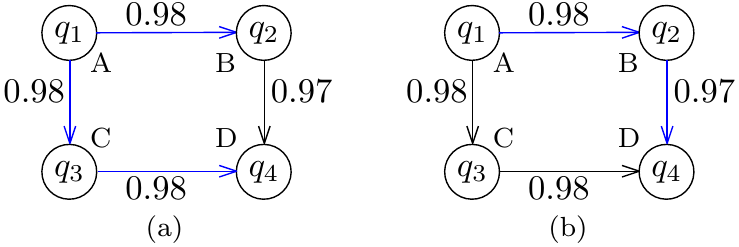}\vspace{-5pt}
    \caption{The data flow (blue paths) of (intra-cluster) multicasting $q_1$ to node B, D: (a) by the shortest path; (b) by the minimum spanning tree. $q_1, q_2, q_3, q_4$ are qubits in node A, B, C, D, respectively. Numbers on edges denote the fidelity of the related EPR pairs.}
    \label{fig:intraexam}
\end{figure}

\paragraph{Intra-cluster communication routing}
For optimizing intra-cluster communication, we should reduce the overall communication footprint of the collective operation, instead of simply shortening paths between any two communication endpoints.
To demonstrate the difference between these two optimization targets, 
we consider the example in Figure~\ref{fig:intraexam} where we are multicasting $q_1$ to nodes B, D. 
The shortest path for establishing the cat-state on $q_1$ and $q_4$ is by $q_1 \to q_3 \to q_4$ as it has the highest overall fidelity along the path. However, this shortest path does not lead to the minimal overall fidelity of the multicast operation. 
As shown in Figure~\ref{fig:intraexam}, using the blue paths $q_1 \to q_2 \to q_4$ in (b), we can establish the cat-state on $\{q_1, q_2, q_4\}$ with only two EPR pairs. In contrast, using the paths $q_1 \to q_3 \to q_4$ and $q_1 \to q_2$ in (a) would need only one more EPR pair to build the cat-state on $\{q_1, q_2, q_4\}$. Moreover, the fidelity of the paths in (a) is 0.98*0.98*0.98 which is worse than 0.98*0.97, i.e., the fidelity of paths in (b).

Thus, using intermediate nodes in the collective communication operation for relaying the communication can greatly improve the fidelity and latency of the collective communication. Specifically, we can optimize the collective communication by finding the minimum spanning tree (MST) of involved qubits, which can be efficiently constructed by Kruskal's algorithms~\cite{kruskal1956shortest}.
Besides being used for building the MST, we can exploit the communication buffer to enable concurrent (collective) communication, in order to reduce the overall communication latency.



\paragraph{Inter-cluster communication routing}

\begin{figure}[t]
    \centering
    \includegraphics[width=0.45\textwidth]{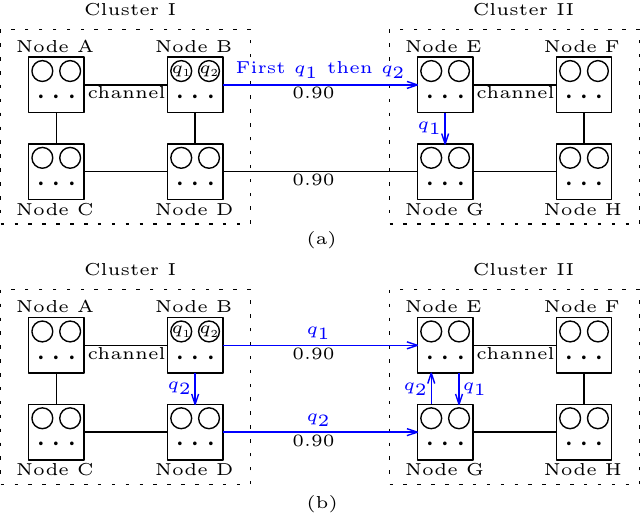}
    \caption{Multicasting $q_1$ to node E, G, and $q_2$ to node E: (a) by the minimum spanning tree; (b) by maximizing the inter-cluster parallelism. Assuming only one currently available communication channel between node B and node E. }
    \label{fig:interexam}
\end{figure}

The optimization of inter-cluster communications is different from optimizing intra-cluster communications as it is much harder and more time-consuming to establish a communication channel between nodes in a different cluster than nodes in the same cluster. 

On the one hand, similar to the MST solution in intra-cluster communication routing, when sharing the qubit $q_1$ in Figure~\ref{fig:interexam} to nodes E, G in the other cluster, we first communicate $q_1$ to node E and use the EPR pair between node E and node G for sharing $q_1$ to node G. If we use the path $q_1 \to D \to G$, we would need another inter-cluster EPR pair which would harm both the fidelity and latency of the collective communication being routed. In summary, when communicating one qubit from one cluster to another cluster, we should use one node in  another cluster as the communication relay to reduce the inter-cluster EPR pairs needed. 

On the other hand, we cannot simply apply the intra-cluster communication routing to the inter-cluster case. For example in Figure~\ref{fig:interexam}, we are multicasting $q_1$ to nodes E, G and $q_2$ to node E, concurrently. There are two ways to fulfill these inter-cluster communications, as shown in Figure~\ref{fig:interexam}(a)(b). The communication path in Figure~\ref{fig:interexam}(a) is simply generated by solving the MST problem. 
The shortcoming of this communication routing is that
we have to wait for a long inter-cluster EPR preparation time for multicasting $q_2$ to node E
as the only available communication channel between node B and E is consumed previously by communicating $q_1$.
To reduce the overall communication latency and improve the information throughput, we propose to use the communication routing shown in Figure~\ref{fig:interexam}(b). That is, we should exploit all available inter-cluster channels for data transfer, instead of sticking to the routing generated by the intra-cluster communication optimization. 
The reason for this routing design is that communication channels between intra-cluster nodes are easier to prepare (thus relatively abundant) while the number of inter-cluster channels is very limited and takes an extremely long time to prepare. For inter-cluster communication, unnecessary stalling would greatly degrade the communication throughput and lead to very high decoherence error.

\section{
    Buffer-aware Distributed Compilation
}~\label{sec:pipeline}
In this section, we assemble compilation components to form the pipeline for generating communication-efficient distributed quantum programs, and discuss the optimization of the communication buffer.

\begin{figure}[t]
    \centering
    \includegraphics[width=0.48\textwidth]{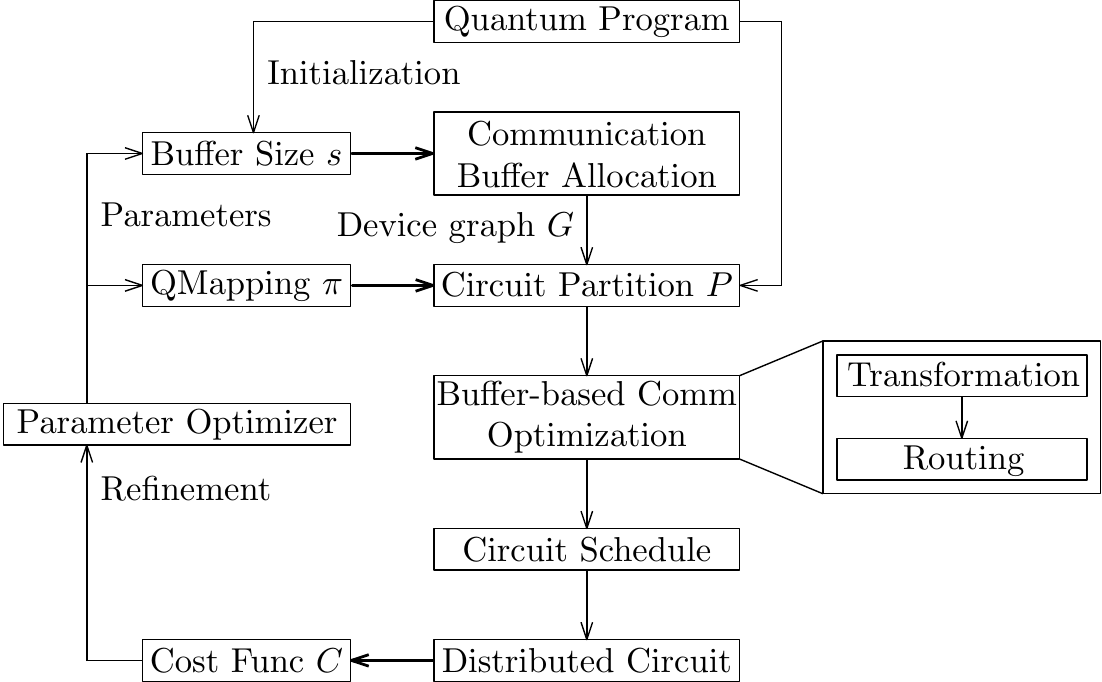}
    \caption{The communication buffer-based compilation pipeline for distributed quantum programs. The compiled distributed circuit is further improved by the parameter optimizer.}\label{fig:pipeline}
\end{figure}

\subsection{Compilation Pipeline Overview}

In the compilation pipeline shown in Figure~\ref{fig:pipeline}, we consider the following basic components:
\begin{enumerate}
    \item Communication buffer allocation, which produces buffer-aware device graph $G$ by reserving some data qubits in each node as the communication buffer. The size of communication buffer $s = (s_1, s_2, \cdots, s_i,\cdots)$ on each node $n_i$ is a tunable parameter;
    \item Distributed circuit partition algorithm $P$, which takes a quantum circuit $circ$ and $G$ as inputs, and generates a distributed version of $circ$ on $G$. The qubit mapping $\pi$ of $P$ is a tunable parameter;
    \item The buffer-based collective communication optimization pass $coll\_comm\_opt\_pass$ proposed in Section~\ref{sec:commopt};
    \item The circuit scheduling pass, which generates the  schedule for operations in the distributed circuit optimized by $coll\_comm\_opt\_pass$. Specifically, for EPR preparation, we would keep establishing EPR pairs among computing nodes for inter-node blocks going to happen and buffer established EPR pairs in the communication buffer for future use. The timing for EPR preparation is adjusted to minimize program stall and EPR pair decaying (due to decoherence).
    \item Cost function $C$, which takes the distributed circuit $dcirc$ generated by the scheduling pass as input, and outputs the estimated program fidelity based on statistics of $dcirc$, e.g., the total number of EPR pairs needed, the overall distributed circuit latency.
\end{enumerate}
Specifically, we define the cost model as follows:
\begin{equation}\label{equ:cost}
    C = f_{1q}^{N_{1q}}f_{2q}^{N_{2q}}f_{ms}^{N_{ms}}f_{iep}^{N_{iep}}f_{oep}^{N_{oep}}(1 - e^{-\frac{t}{T}})
\end{equation}
where $f_{1q}$, $f_{2q}$, $f_{ms}$, $f_{iep}$, $f_{oep}$ are defined in Table~\ref{tab:netdata}, $N_{1q}$, $N_{2q}$, $N_{ms}$, $N_{iep}$, $N_{oep}$ are counts of the operations, $t$ is the program latency, and $T$ is the coefficient for decoherence error.

The two tunable parameters $s$ (communication buffer size on compute nodes) and $\pi$ (qubit mapping) in the compilation pipeline affect the communication efficiency of the compiled distributed program, and we adopt the simulated annealing algorithm to optimize them, with $-\log C$ as the energy function, $s$ being optimized first and $\pi$ the second. 
The initial values of $\pi$ and $s$ greatly affect the iteration cost of simulated annealing and the values found. 
We initialize $\pi$ with the commonly-used graph partition algorithm-- Overall Extreme Exchange (OEE) by Park and Lee~\cite{Park1995AlgorithmsFP}. This algorithm can be used to find the qubit mapping that minimizes the interaction between compute nodes by exchanging qubits. To ease the design of the communication buffer, we preserve one data qubit in each node. The generated qubit mapping serves as a good initialization as it provides a lower up bound for communication overhead. 

In the next section, we discuss the initialization of the buffer parameter $s$, which plays a critical role in the proposed buffer-aware compilation pipeline.

\subsection{Communication Buffer Design}


\begin{figure}[t]
    \centering
    \includegraphics[width=0.35\textwidth]{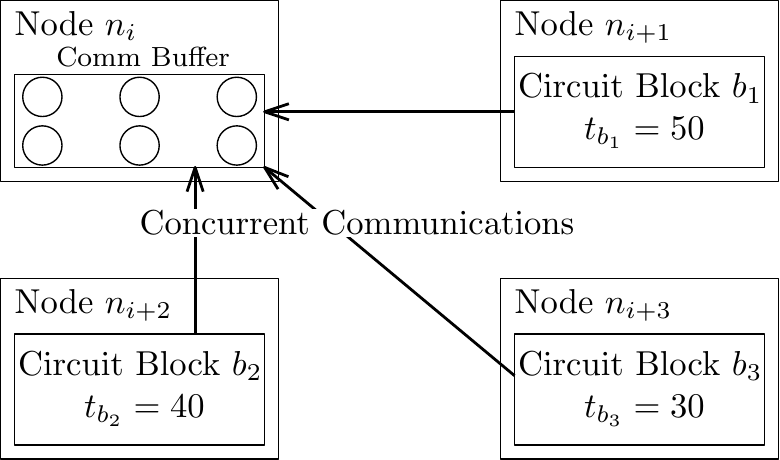}
    \caption{Communication buffer size initialization example. Node $n_{i+1}$ is in the same cluster as the node $n_i$.}
    \label{fig:buffersize}
\end{figure}

For communication buffer initialization, we consider the distributed circuit mapped by the initial $\pi$ and optimized by the burst communication technique proposed in~\cite{autocomm}. 
For an arbitrary non-idle node $n_i$ (i.e., $\exists$ program qubits mapped into $n_i$), we compute initial $s_i$ as follows:
\begin{equation}\label{equ:bufsize}
    s_i = 2*\max(\max_B \lfloor \frac{t_B}{t^B_{ep}+t_{buf}} \rfloor, R_{pb}(n_i))
\end{equation}
Here $B$ is an inter-node block (involving two compute nodes) with $n_i$ being the communication target node, i.e., the block $B$ would be locally computed in $n_i$ with related quantum data shared/teleported to $n_i$. $t^B_{ep}$ is the latency for preparing EPR pairs between the endpoints of $B$. If the endpoints of $B$ are in the same cluster as $n_i$, then $t^B_{ep} = t_{iep}$, otherwise, $t^B_{ep} = t_{oep}$. $t_{buf}$ is the latency of the selected EPR buffering protocol, e.g. $t_{buf}=2$ for the one in Figure~\ref{fig:virtualcomm}(b). $R_{pb}(n_i)$ is the maximum number of parallel inter-node blocks involving $n_i$. 
The `2*' in Equ~(\ref{equ:bufsize}) means that we would use half of the communication buffer in each node for handling communication requests, and the other half for buffering EPR pairs. 
To give a concrete example of computing $s_i$, we consider the example in Figure~\ref{fig:buffersize}. Since there are three concurrent inter-node blocks involving $n_i$, we have $R_{pb}(n_i) = 3$. On the other hand, $\max_B \lfloor \frac{t_B}{t^B_{ep}+t_{buf}} \rfloor = \lfloor \frac{50}{12+2}\rfloor = 3$ (note that block $b_1$ is the largest remote block involving $n_i$). Therefore, for this example, we just $s_i = 2*3 = 6$.

This communication buffer design 
could achieve the maximal EPR generation rate of $n_i$, i.e., the communication qubit of $n_i$ is not stalled by a specific lengthy communication request while keeping $n_i$ always ready for concurrent communication requests. On the other hand, a large communication buffer resulting from $s_i$ can accommodate more aggressive collective communication optimization discussed in Section~\ref{sec:commopt}. 

In practice, it is possible that the sum $\sum_i s_i$ is larger than the number of idle qubits in non-idle nodes, say $N_{idleq}$. In such a case, we adjust $s_i$ proportionally to make $\sum_i s_i \le N_{idleq}$.
This constraint on the communication buffer size $s$ is also enforced when we pick the neighborhood of $s$ in simulated annealing, which is adopted to achieve a good trade-off between program qubits and virtual communication qubits in each node.



\section{Evaluation}\label{sec:eval}

In this section, we compare the performance of {\frameworkName} to the baseline~\cite{autocomm} and analyze the effect of optimization passes proposed in {\frameworkName}. 

\subsection{Experiment Setup}\label{sect:expset}

\paragraph{Platforms} 
All experiments are executed on a Ubuntu 20.04 server which is equipped with one 28-core Intel Xeon Platinum 8280 processor and 1TB RAM.
The used software includes Python 3.9.13 and Qiskit 0.37.0~\cite{Qiskit}.

\paragraph{Benchmark programs} The benchmark programs used in the evaluation are summarized in Table~\ref{tab:benchmark} and are derived from IBM Qiskit~\cite{Qiskit} and RevLib~\cite{revlib}. Programs in Table~\ref{tab:benchmark} are divided into two categories. The first category consists of building blocks of quantum applications, including multi-controlled gate (MCTR), quantum ripple carry adder (RCA) and cat-state preparation (CSP). The benchmarks in the first category are small circuits.
The second category contains relatively large programs that represent various quantum applications, e.g., quantum Fourier transformation (QFT), Bernstein-Vazirani (BV) algorithm, and Quantum Approximate Optimization algorithm (QAOA). For BV, we consider 1000 randomized secret strings which each contain $0.7*\#\,qubit$ nonzeros. For QAOA, we consider the maxcut problem on graphs where each vertex connects $30+\lfloor \frac{\# qubit}{12} \rfloor$ vertices on average.

\paragraph{Baseline} As the baseline, we implement the DQC compiler AutoComm~\cite{autocomm}. AutoComm
groups a series of remote CX gates between two compute nodes into a \textit{burst} communication block and implements the burst communication block (all remote CX gates in it) with at most two EPR pairs. AutoComm represents currently the best effort in optimizing quantum communication overhead in distributing quantum programs, as far as we know. However, without the communication buffer, AutoComm cannot optimize more general collective communication that may involve more than two compute nodes. In the following evaluation, we adopt the same circuit partition algorithm---OEE algorithm (ref. Section~\ref{sec:pipeline}) for both {\frameworkName} and AutoComm, in order to eliminate the difference caused by the circuit partition.

\paragraph{DQC network model}   
For evaluation, we adopt the DQC computing system in Figure~\ref{fig:netmodel}. Specifically, we assume two clusters in the DQC network, with each cluster containing four compute nodes and each compute node containing 40 data qubits and 1 communication qubit. The EPR pairs/channels that can be directly established between two compute nodes are depicted in Figure~\ref{fig:netmodel}. The inter-cluster EPR pair and intra-cluster EPR pair differ from each other significantly in terms of fidelity and preparation time, as shown in Table~\ref{tab:netdata}. Like in the baseline, to focus on communication optimization,
we assume trapped-ion style compute nodes~\cite{bruzewicz2019trapped} where any two local qubits can communicate with each other directly. Note that our work can be easily extended to superconducting compute nodes by integrating with existing single-node routing passes~\cite{Qiskit}.

\paragraph{Metric} To characterize the communication overhead of the compiled distributed circuits, we consider the following metrics:
\begin{enumerate}
    \item The number of consumed EPR pairs (`\# EPR'). More EPR pairs consumed would naturally lead to lower program fidelity and longer latency. Thus, a lower value of `\# EPR' is preferred.
    \item The number of consumed inter-cluster EPR pairs (`\# CROSS EPR'). `\# CROSS EPR' should be minimized due to the low fidelity of inter-cluster EPR pairs and the long latency to prepare them.
    \item The latency of the compiled distributed circuit (normalized to CX counts). 
    Longer latency would lead to higher decoherence error, thus a lower value is preferred.
    \item The size of allocated communication buffers to model the resource consumption induced by using communication buffers. Specifically, we consider the average number of qubits in communication buffers on compute nodes (`AVG CB \#q'). 
\end{enumerate}
To model the relative performance of {\frameworkName} to the baseline, we consider the following metrics (expected as large as possible):
\begin{enumerate}
    \item `EPR-DEC (\%)', which is defined by `1 - (\# EPR by {\frameworkName} / \# EPR by baseline)*100\%'.
    \item `CROSS-EPR-DEC (\%)', which is defined by `1 - (\# CROSS EPR by {\frameworkName} / \# CROSS EPR by baseline)*100\%'.
    \item `LAT-DEC (\%)', which is defined by `1 - (latency by {\frameworkName} / latency by baseline)*100\%'.
    \item `FD-INC (\%)', which is defined `(ES-FD by {\frameworkName} / ES-FD by baseline - 1)*100\%'. The `ES-FD' (estimated fidelity) is computed according to Equ~(\ref{equ:cost}).
\end{enumerate}


\begin{table*}[t]
\centering
\caption{The compilation results of {\frameworkName} and its relative performance to the baseline on benchmark programs. 
}
\label{tab:benchmark}
\def\nsize {1.75cm}
\resizebox{0.91\textwidth}{!}{
\renewcommand*{\arraystretch}{1.3}
\begin{tabular}{|p{1.0cm}|p{\nsize}|p{0.7cm}|p{0.8cm}|p{0.8cm}|p{0.8cm}|p{1.2cm}|p{0.9cm}|p{0.8cm}|p{1.2cm}|p{0.8cm}|p{1.cm}|} 
    \hline
    \multirow{2}{*}{Type} & \multirow{2}{*}{Name} & \multirow{2}{*}{\# qubit} & \multicolumn{5}{c|}{\frameworkName} & \multicolumn{4}{c|}{Compared to AutoComm} \\ 
    \cline{4-12}
     &  &  & \# node & AVG CB \#q & \# EPR & \# CROSS EPR & Latency & EPR-DEC & CROSS-EPR-DEC & LAT-DEC & FD-INC \\ 
    \hline
    \multirow{9}{1cm}{Building Blocks} & \multirow{3}{\nsize}{Multi-Controlled Gate (MCTR)} & 100 & 3 & 1.333 & 2 & 0 & 2521.3 & 0.6 & 0 & 0.045 & 0.109 \\ 
    \cline{3-12}
     &  & 200 & 6 & 1.667 & 5 & 2 & 2547.7 & 0.8 & 0.333 & 0.263 & 1.195 \\ 
    \cline{3-12}
     &  & 300 & 8 & 1.75 & 7 & 4 & 2917.3 & 0.811 & 0.636 & 0.398 & 4.499 \\ 
    \cline{2-12}
     & \multirow{3}{\nsize}{Ripple-Carry Adder (RCA)} & 100 & 3 & 1.333 & 4 & 0 & 872.6 & 0.5 & 0 & 0.081 & 0.139 \\ 
    \cline{3-12}
     &  & 200 & 6 & 1.667 & 10 & 2 & 1775.8 & 0.375 & 0.5 & 0.151 & 0.470 \\ 
    \cline{3-12}
     &  & 300 & 8 & 1.75 & 14 & 2 & 2664.6 & 0.5 & 0.5 & 0.143 & 0.900 \\ 
    \cline{2-12}
     & \multirow{3}{\nsize}{Cat-State Preparation (CSP)} & 100 & 3 & 0.333 & 2 & 0 & 63.4 & 0 & 0 & 0.427 & 0.005 \\ 
    \cline{3-12}
     &  & 200 & 6 & 0.667 & 5 & 1 & 158.4 & -0.25 & 0 & 0.547 & -0.011 \\ 
    \cline{3-12}
     &  & 300 & 8 & 0.75 & 7 & 1 & 164.4 & 0 & 0.75 & 0.792 & 0.374 \\ 
    \hline
    \multirow{9}{1cm}{Real World Appli-cations} & \multirow{3}{\nsize}{Quantum Fourier Transformation (QFT)} & 100 & 3 & 6 & 98 & 0 & 3903.9 & -0.225 & 0 & 0.638 & 0.148 \\ 
    \cline{3-12}
     &  & 200 & 6 & 6 & 492 & 64 & 9790.5 & -0.23 & 0.6 & 0.772 & $>>1e^3$ \\ 
    \cline{3-12}
     &  & 300 & 8 & 2.5 & 1058 & 152 & 38822.2 & -0.080 & 0.729 & 0.913 & $>>1e^9$ \\ 
    \cline{2-12}
     & \multirow{3}{\nsize}{Bernstein Vazirani (BV)} & 100 & 3 & 0 & 1 & 0 & 55.3 & 0 & 0 & 0.002 & 0.000 \\ 
    \cline{3-12}
     &  & 200 & 5 & 0.333 & 3 & 0 & 73.4 & 0 & 0 & 0.583 & 0.010 \\ 
    \cline{3-12}
     &  & 300 & 8 & 0.5 & 5 & 1 & 160.4 & 0 & 0.5 & 0.661 & 0.123 \\ 
    \cline{2-12}
     & \multirow{3}{\nsize}{QAOA} & 100 & 3 & 4 & 60 & 0 & 1323.0 & 0 & 0 & 0.476 & 0.128 \\ 
    \cline{3-12}
     &  & 200 & 6 & 4 & 300 & 40 & 3545.0 & 0 & 0.75 & 0.726 & $>>1e^3$ \\ 
    \cline{3-12}
     &  & 300 & 8 & 2.5 & 560 & 80 & 8294.0 & 0 & 0.75 & 0.958 & $>>1e^9$ \\
    \hline
\end{tabular}
}\vspace{-10pt}
\end{table*}

\subsection{Compared to the baseline}

In this section, we analyze the relative performance of {\frameworkName} compared to the baseline and discuss the effect of each optimization pass in \frameworkName. Table~\ref{tab:benchmark} summarizes the relative performance of {\frameworkName} compared to the baseline.

Firstly, compared to the baseline, {\frameworkName} on average reduces `\# EPR' (the total number of EPR pairs consumed) by 14.6\%, 11.6\%, and 20.5\% on 100-qubit, 200-qubit, 300-qubit programs, respectively.
The optimization of `\# EPR' in {\frameworkName} is affected by two factors: the communication buffer and the communication transformation pass. The communication buffer alone has a negative impact on reducing `\# EPR'. This is because the communication buffer may lead to more occupied nodes and more scattered qubit layouts when distributing the circuit and thus more inter-node communications. For example, for the 200-qubit CSP program, {\frameworkName} requires 6 nodes while AutoComm requires 5 nodes. The one more node usage leads to one more EPR pair by {\frameworkName}. 

On the other hand, the communication transformation pass could reduce the EPR pair consumption remarkably. 
For example, for RCA, the major communication is TP-Comm based.
`\# EPR' used by {\frameworkName} is smaller due to the lazy teleportation technique in the communication transformation pass which utilizes the communication buffer to temporarily store the teleported qubit instead of immediately moving the teleported qubit back to its original node as in AutoComm. For MCTR, {\frameworkName} largely reduces `\# EPR' (73.0\% on average) because of the communication fusion technique which can optimize multi-party communication using the communication buffer. The MCTR by AutoComm is already optimized by the communication splitting technique in {\frameworkName}. Otherwise, to deal with the multi-controlled gate, AutoComm has to decompose it into the CX+U3 basis~\cite{Qiskit} which would incur thousands of EPR pair consumption, making AutoComm even worse compared to {\frameworkName}.

Secondly, the communication routing pass in {\frameworkName} greatly reduces inter-cluster communication costs. Except for 100-qubit programs which do not involve inter-cluster communication, {\frameworkName} reduces `\# CROSS EPR' (the total number of inter-cluster EPR consumed) on average by 36.4\% and 64.4\%, on 200-qubit and 300-qubit programs respectively. The reduction of `\# CROSS EPR' on 200-qubit programs is smaller than that on 300-qubit programs because inter-cluster communication is relatively limited on 200-qubit programs.
Overall, we observe that
the significant reduction of {\frameworkName} on inter-cluster EPR pairs owes to the optimization in communication routing pass that converts inter-cluster communications into inter-node communications by using one compute node in the (communication) target cluster for communication relaying. To illustrate this point clearly, though the `\# EPR' by {\frameworkName} for CSP, QFT, BV, and QAOA is more than (or equal to) that by AutoComm, `\# CROSS EPR' by {\frameworkName} on these programs is still considerably lower than that by AutoComm.

Moreover, {\frameworkName} significantly reduces the latency of distributed programs. Compared to the baseline, {\frameworkName} reduces the program latency on average by 27.8\%, 50.7\%, and 64.4\% on 100-qubit, 200-qubit, and 300-qubit programs, respectively. For 100-qubit programs, there is no inter-cluster communication and the latency reduction is mainly due to the high communication parallelism/throughput enabled by a large number of virtual communication qubits in communication buffers. The latency reduction by {\frameworkName} grows as inter-cluster communication increases due to the great ability of {\frameworkName} in optimizing inter-cluster communication.

Not surprisingly, {\frameworkName} also boosts the fidelity of compiled programs greatly. 
For 100-qubit programs where the number of inter-node communication is relatively limited, {\frameworkName} still achieves an 8.8\% fidelity improvement over AutoComm on average. For larger programs, the fidelity improvement by {\frameworkName} is even larger. Except for QFT and QAOA, {\frameworkName} reduces the program latency on average by 41.6\% and 147.4\% on 200-qubit and 300-qubit programs, respectively compared to AutoComm. For QFT and QAOA, {\frameworkName} provides overwhelming fidelity improvement due to the significantly reduced inter-cluster EPR pairs and program latency.

In summary, {\frameworkName} is obviously better than state-of-the-art DQC compilers in optimizing collective communication on the hierarchical and heterogeneous DQC network.

\section{Related Work} 

\paragraph{Compilers for monolithic QC:} Previous quantum compilers~\cite{Qiskit, Li2019TacklingTQ, Amy2019staqAFQ, Sivarajah2020tketAR, Khammassi2022OpenQLA} compile quantum programs to a monolithic QC hardware. While we can extend these works to DQC by treating the transformation and routing remote communications like local/in-node communications, this simple extension cannot fully expose the computational potential of DQC. This is because, unlike this paper, these compilers do not optimize collective communications widely existing in distributed quantum programs.

\paragraph{Compilers for DQC:} Compilers for DQC can be mainly divided into two categories. Works in the first category~
\cite{Beals2013EfficientDQ, Moghadam2016OptimizingTC, AndresMartinez2019AutomatedDO, Davarzani2020ADP, daei2020optimized, time-slice, Hner2021DistributedQC, Dadkhah2022ReorderingAP} ignore the low-level implementation of quantum communication and perform program optimizations on the logical level, equivalent to assuming unlimited communication qubits. These works mostly focus on the circuit partition and do not consider advanced  transformation and routing strategies for collective communication as efficient communication is naturally enabled by the unbounded communication resource. Thus we cannot guarantee the performance of these works on a communication-resource-constrained DQC system. While 
Häner et al.~\cite{Hner2021DistributedQC} provided optimizations for broadcast and reduce operations in DQC, their work optimizes these collective operations as library functions and does not consider the program context and the underlying hardware topology, which are essential for compiler optimizations in this paper.

Works in the second category~\cite{Ferrari2021CompilerDF, Diadamo2021DistributedQC, autocomm} consider communication optimization on DQC hardware with limited communication qubits as in this paper. However, these works fall short in implementing multi-node operations due to the shortage of communication qubits, let alone exploring the transformation and routing of collective communication. To execute multi-node operations, these works would decompose them into small multi-node circuit blocks which would inevitably incur higher communication costs. 
Besides the inefficiency in performing collective communication, these works admit low communication throughput. They usually perform inter-node communications with quite limited parallelism and even sequentially, due to the lack of efficient utilization of the constrained communication resource. All these drawbacks make these works less efficient than this paper in optimizing collective communication.

Finally, there are also compilers~\cite{cutQC, Peng2020SimulatingLQ} proposed to execute distributed quantum programs in a way without remote communications. These works cut large circuits into smaller ones and then run these small circuits on different devices. 
The drawback of these works is that they require unscalable classical post-processing for large-scale DQC. Instead, our paper does not rely on classical post-processing and can be scaled to the large-scale DQC.



\section{Discussion and Future Work}

To the best of our knowledge, this paper is the first attempt that formalizes a concept of the communication buffer in DQC.
The communication buffer leads to a new computing model lying between the one on a logical DQC network and the other one on a physical DQC system and unveils new optimization opportunities for collective communication in DQC. We believe this paper would open new research topics for  communication optimization in DQC.

Although we show that the proposed framework significantly surpasses existing works in optimizing the collective communication of distributed quantum programs, there is still much space left for potential future works. 

\paragraph{Exploring more usage of communication buffer}
This paper mainly focuses on the benefit of the communication buffer in communication transformation and routing. There is actually more usage of the communication buffer. For example, we can also use the communication buffer for distilling the EPR pairs, especially the inter-cluster EPR pairs. 
Optimizing collective communication along with EPR pair distillation presents new research opportunities. Besides, the communication buffer can also be used to implement more complicated communication protocols beyond the Cat-Comm and TP-Comm. For example, we may replace a physical qubit in quantum communication with a logical qubit so that the logical qubit serves as a bridge to two compute nodes with communication channels hidden from programming and mainly used for the error detection on the logical qubit.

\paragraph{Extending to fault-tolerant DQC program}
This paper only considers the collective communication of unprotected distributed quantum programs. For future DQC, it is expected programs will be built upon quantum error correction (QEC) codes to account for the noise from inter-node communications. If we consider the fault-tolerant architecture~\cite{Jnane2022MulticoreQC} which constructs one logical qubit upon multiple nodes, basic QEC operations such as magic state distillation, logical gates, and stabilizer measurements will all inevitably incur collective communication, making the fault-tolerant DQC program more demanding for collective communication optimization. It's thus interesting to extend our work to optimize fault-tolerant DQC programs.


\paragraph{Adapting to more DQC architectures} 
The proposed framework mainly explores the network hierarchy of DQC. Considering recent progress in constructing quantum networks~\cite{Hermans2022QubitTB,QNexp9,QNexp10,QNexp11,QNexp12,QNexp13,QNexp14,QNexp15,QNexp16,QNexp17,QNexp18,QNexp19}, we can inspect more hardware levels in DQC architecture. For example, a DQC system may consist of multiple physically distinguished kinds of qubits~\cite{Hermans2022QubitTB}: memory qubits, communication qubits, etc. We can use memory qubits that have a long coherence time to build the communication buffer and use the native operations on the memory qubit as the new instruction set on the communication buffer to construct QEC in the communication buffer or to optimize collective communication. For program qubits, we may adopt a different type of qubits from memory qubits, e.g. superconducting qubits which are relatively accurate for quantum computation. In such a case, the communication between the communication buffer and program qubits becomes a new optimization problem to explore.



\section{Conclusion}

To promote the computational power of DQC, we invent the \textit{communication buffer}
to decouple the execution of inter-node operations from communication qubits.
This design paves the way for collective communication optimization. We then propose \frameworkName, a buffer-based compiler framework targeting the optimization of collective communication in distributed quantum programs.
Experimental results on a hierarchical DQC system show that the proposed \frameworkNameSpace can reduce the most expensive inter-node communication request and the program latency by 50.4\% and 47.6\% on average, respectively.

\section*{Acknowledgment}
This material is based upon work supported by the U.S. Department of Energy, Office of Science, National Quantum Information Science Research Centers, Quantum Science Center. The Pacific Northwest National Laboratory is operated by Battelle for the U.S. Department of Energy under Contract DE-AC05-76RL01830. This work was also supported in part by NSF 2048144.


\bibliographystyle{unsrt}
\bibliography{references}

\end{document}